\documentstyle[sprocl,twoside,axodraw,amsmath,amssymb]{article}

\def\Journal#1#2#3#4{{#1} {\bf #2}, #3 (#4)}
\def\NCA{\em Nuovo Cimento}

\def\NPB{{\em Nucl.~Phys.} B}
\def\PLB{{\em Phys.~Lett.} B}

\def\Ann{\em Ann.~Phys.}
\def\YAF{\em Yad.~Fiz.}
\def\SJNP{\em Sov.~J.~Nucl.~Phys.}
\def\ZhETF{\em Zh.~Exp.~Theor.~Fiz.}
\def\JETP{\em Sov.~Phys.~JETP}
\def\Acta Pol{{\em Acta~Phys.~Pol.} B}

\def\be{\begin{equation}}
\def\ee{\end{equation}}
\def\bea{\begin{eqnarray}}
\def\eea{\end{eqnarray}}
\def\nn{\nonumber}
\def\d{\partial}
\def\b{{\bf b}}
\def\k{{\bf k}}
\def\z{{\bf z}}
\def\q{{\bf q}}
\def\u{{\bf u}}
\def\B{{\cal B}}

\def\K{{\cal K}}
\def\Z{{\cal Z}}
\def\Q{{\cal Q}}
\def\II{{\cal I}}
\def\F{{\cal F}}
\def\G{{\cal G}}
\def\y{\tilde{y}}
\def\h{\tilde{h}}
\def\f{\tilde{f}}
\def\w{\tilde{w}}
\def\g{\tilde{g}}
\def\s{\tilde{s}}

\def\T{\hat{T}}
\def\H{\hat{H}}
\def\p+{\! + \!}
\def\m-{\! - \!}
\def\r={\! = \!}
\def\Im{{\mathrm Im}}
\def\ds{\displaystyle}
\def\scs{\scriptstyle}
\def\scss{\scriptscriptstyle}
\def\N{{\scss N}}
\def\A{{\scss A}}
\newcommand{\bmath}[1]{\mbox{\boldmath #1}}

\allowdisplaybreaks
\begin{document}
\sloppy
\title{PROBABILISTIC MODEL OF REGGEON FIELD THEORY}
\author{K. G. BORESKOV}
\address{
Institute of Theoretical and Experimental Physics,
Moscow, Russia\\E-mail: boreskov@heron.itep.ru}

\maketitle\abstracts{
The stochastic model of classical system of particles (partons),
which dynamics includes random walk in plane as well as processes
of death, splitting, annihilation and fusion of partons, is considered.
A set of equations for multiparticle distribution functions
for this system can be described in terms of diagrams of the Reggeon 
Field Theory (RFT) with supercritical pomeron, where rapidity plays 
a role of the time variable. The multiparticle inclusive distributions 
of partons correspond to multi-pomeron states in this analogy.
In order to calculate the amplitude of hadron interaction 
at given energy we define an operator of linkage of any pair of partons
from two parton sets (the projectile and the target). 
The form of the hadron-interaction (parton-linkage) operator is 
determined from the requirement for the result to be in correspondence
to the RFT formulas. It is shown that the requirement of Lorentz 
invariance of the amplitude is fulfilled in this interpretation only 
at definite relation between the probabilities of parton splitting, 
annihilation and fusion. 
Interpretation of various methods of $s$-channel unitarization
is discussed in terms of the model.}

\tableofcontents
\newpage
\section*{~}
\centerline{***}
{\small
\hspace*{5.7cm} Yet say not sadly: they have left us! \\
\hspace*{5.7cm} But say, with gratitude: they were."\\
\hspace*{9cm}--- Vasily Zhukovsky \\
}

This paper is dedicated to memory of Misha Marinov. One of the
first papers published by Misha was a work done together with
Roginsky on relation between $s$- and $t$-channel
helicity amplitudes which was important for description of spin
particle interactions in Regge theory. I've just come to this field
and was strongly impressed
with mathematical elegance of this paper, especially with the fact
that the direction to the unphysical point (centre-of-mass of $t$
channel) turned out to be real and physical. Unfortunately, this paper
contained a minor mistake (change of helicity sign for transition
from particle to antiparticle was not taken into account). I
think that this mistake was related with Misha's style because it
could be easily detected by consideration of the simplest example
of $\pi N$ scattering. But Misha liked general considerations so
much that didn't carry on elementary test. Later on these results
were independently reproduced (without the mistake) in the well
known paper by Trueman and Wick where examples were carefully
considered.

In Theoretical Department Misha acted as a receptor providing an information about
new physical ideas especially using advanced mathematical technique. He was in earnest
about this mission and prepared these talks very carefully. He always had a detailed plan
which he followed very hard and didn't allow uncontrolled discussions and digressions
if they were not planned. Due to this hardness he always managed to convey to the audience
main ideas
and his own relation to them. Up to now my attitude to many of these topics
is affected by primary Misha's interpretation.

Misha was born Teacher, and it was a mystery for me why he had no students in ITEP
to teach or to supervise them. Once I asked him to give a couple of popular lectures
on path integral for me and some colleagues. He agreed instantly, and as a result we got
a half-year course of lectures, which was transformed eventually to the remarkable review
in Physics Reports. And how many such opportunities we missed!

Everything what he did, Misha tried to execute professionally,
either it was scientific paper writing, tea making, or karate training.
I remember as during one of ITEP Winter Schools we were playing
charades trying to show meaning of word or  its part by
performing short sketches. Misha chose to show just the word
"charade" (in russian). First his team performed three sketches
for its parts, and then, to show the word as a whole, they
repeated all these sketches again at a rapid pace, representing
the process of the game. It was well done, and we could see
perfectionism of Misha as a theater director. To be more
professional as an actor he used tooth-paste for his make-up
(performing a role of devil).

Misha's unlimited erudition and knowledge were always at service of his colleagues.
They consulted with him on any subject and always got thoughtful discussion and
useful references, often to old and forgotten papers. The paper which I present in this
volume just illustrates the lack of Misha's advices and consultations for me -- when
it has been finished I met the old paper by Grassberger and Sundermeyer where the same
model for reggeon field theory was formulated
\footnote{~I am grateful to A.~Kaidalov and M.~Ryskin who drawed my attention to this paper.}.

I liked to discuss with Misha Jewish history, origin of Jewish names, problems of Jewish
young people at entrance examinations in universities, etc.
But not less amount of exciting information I got from him on Russian history
and language, literature, biology, and many other things (by the way, the origin of the
word "eikonal" appearing in this paper and common with the word "icon" I knew from Misha).

Misha's high standards exerted influence on attitude to different situations, including
ethical problems, in the circle of ITEP theorists. His opinion often was crucial giving,
as a tuning fork, a point of reference. And now, what unphysical point Misha would be at,
the orientation to his standards is still fixed for us.
\section{Introduction}
Processes of strong interactions at high energies are described usually
in framework of reggeon theory \cite{reprints}.
Systematic description of high--energy interactions of hadrons in terms
of reggeon field theory (RFT) was introduced by Gribov in ref.~\cite{gribov67}.
These results were obtained in a hybrid field theory model by calculation
of Feynman diagrams containing the reggeon amplitudes as building blocks.

Experimental data on hadron interactions show that the total
cross section is dominated by multiparticle processes and the number of secondaries
increases with energy growth. Therefore only models taking into account large number
of effective degrees of freedom (referred below as to {\em partons}
for brevity) in a fast hadron can correspond to the nature of reggeon amplitudes.

Attempts to give the parton interpretation of the reggeon amplitudes started immediately
after appearance of the notion of the reggeon. In ref.~\cite{AFS62} a representation
of the pole reggeon diagram was introduced as a multiperipheral fluctuation
of the fast hadron.
Simultaneous interactions of several multiperipheral fluctuations with the target
correspond to nonplanar reggeon diagrams and explain the presence of the regge-cut
contributions.\cite{mandelstam62}
Later on interpretations of reggeon interactions were suggested using as inner
degrees of freedom the objects of quantum chromodynamics --
the quark--gluon string model~\cite{kaidalov-qsr}, the BFKL pomeron~\cite{BFKL}, etc.

The common feature of all these models is consideration of a fast moving hadron
as a complicated composite system which components (partons) are continuously
distributed in rapidity values and randomly in the impact parameter plane.
Hence, the Fock wave function of the fast hadron is a superposition of multiparton
states and its structure depends on the hadron energy.

Interaction of the hadron with a target
is realized only due to interactions of the most slow (in the target frame) partons
produced step by step from faster components (short-range interaction in the rapidity space).
The amplitude of interaction is determined by the number of slow partons $n(y)$
in the vicinity of the target.
The probability of interaction with a target of only one of slow partons is proportional
to $n(y)$ and corresponds to the regge-pole contribution. Interactions of two and more
partons correspond to the regge cuts of second and higher orders.
The cut contributions are sign-alternating because of shadowing effects.

Growth of hadron cross sections with energy gives evidences in favor of the so called
{\em supercritical} pomeron ($\Delta_P =\alpha_P (0)-1 > 0$).
RFT with the supercritical pomeron encounters a number of difficulties which are not
resolved in satisfactory way. The pole amplitude corresponding to the one-pomeron exchange
increases with energy as $s^{\Delta}$ violating $s$-channel unitarity.
Diagrams of higher orders, in principle, have to restore the unitarity. However,
when the energy increases the number of essential diagrams grows drastically, and
there is no satisfactory way to sum a series of reggeon diagrams at present.
Usually one takes into account only a set of the simplest diagrams corresponding
to the eikonal or ``quasi-eikonal'' approximation
which corresponds to account for the non-enhanced cut only and
disregards inter-pomeron interactions. The scattering amplitude in this approximation
looks as for scattering on almost black disc of radius proportional to the rapidity $y$
in correspondence to the Froissart regime.
In some approximations contributions of  the enhanced regge cuts were summed.
E.g. in the Schwimmer approximation~\cite{Schwimmer75}, which is justified for
hadron--nucleus interactions,
the fan-type enhanced diagrams are taken into account but reggeon loops are neglected.
Analysis of more complicated sets of diagrams was carried out in refs~\cite{Cardy74,DTM,KPTM86}.
However, the problem of construction of hadron scattering amplitude
satisfying both $s$-channel and $t$-channel unitarity is not solved up to now.

In the parton picture the increase of the one-pomeron contribution with energy
in the regime of the supercritical pomeron means that the number of soft partons
grows exponentially with hadron rapidity, $n(y)\sim \exp(\Delta y)$.
Thus, the mechanism of parton multiplication has to be present
in the course of the rapidity evolution of the partonic fluctuation.
It could be splitting of the multiperipheral ladder, gluon cascading, etc.
From general arguments fusion of partons may also take place beside splitting.

The kinetic model of parton interactions equivalent to RFT was suggested
in paper by Grassberger and Sundermeyer \cite{Grassberger78}.
In fact, this model gives a desription of stationary structure of the Fock wave function
of a fast hadron.

In presented paper we discuss interaction of two parton systems in this
model~\footnote{~Unluckily, I found the paper \cite{Grassberger78} only after this text
has been finished. I elaborated the same kinetic model for RFT so the content of
first section overlap considerably with ref.~\cite{Grassberger78}.}.
We analyze a possibility of consistent probabilistic interpretation of the interaction
amplitude in this approach and, particularly, its Lorentz invariance.

The main assumption of the model is continuous parton evolution in rapidity,
i.e. an origin of slower partons from the faster predecessors. In the course
of evolution partons undergo a random walk in the impact parameter plane.
Other essential ingredients of parton dynamics are processes of splitting
of one parton into two ones and fusion of two partons into one parton.
For generality we consider also vanishing (``death'') one of partons at given rapidity and
``annihilation'' of pair of partons at close points.
The correspondence can be proved between multiparticle parton distribution functions
and multipomeron vertices of the RFT with lagrangian including the elementary
three-pomeron coupling and the pomeron-scattering term.
Thus, the pomeron in this model is not some specific object but just one-parton
inclusive distribution, while many-pomeron states are connected to the inclusive
multiparton distributions.

In order to introduce into the model hadron--hadron interactions it is necessary
to define the operator of the interaction of two parton systems with different rapidities.
We show here that this operator can be defined in the way maintaining the correspondence
with the interaction amplitude given by summing all diagrams of RFT including loop diagrams.
It is shown that if some definite relation between the constants of splitting, annihilation and fusion
of partons holds, the hadron interaction amplitude depends only on the sum of rapidities of
the interacting hadrons according to the Lorentz invariance requirement of the RFT.
However the price for the correspondence with RFT is a loss of the probabilistic interpretation,
because the interaction operator doesn't have, in general, a probabilistic meaning. Nevertheless,
the interaction amplitude does not violate $s$-channel unitarity, if initial parton distributions
can be represented as superpositions of Poisson distributions.

The layout of the paper is as follows. In sect.2 the model is formulated in simplified version
which does not take into account parton diffusion in the transverse plane. In sect.3 the diffusion
of partons in the impact parameter plane is included and equations for the multiparticle distribution
functions are derived both in coordinate and momentum representations. The equivalence
of the equations for these multiparton distributions and equations for the many-pomeron vertices
in the ``cut'' RFT is demonstrated in sect.4. In sect.5 we define the interaction operator for two
parton sets and demonstrate independence of inelastic cross section on a choice of Lorentz frame.
In sect.6 the main model approximations in RFT are considered --- the eikonal approximation,
the Schwimmer model and some others.
The picture of an expanding black disc corresponding to the Froissart regime is discussed.
The problem of $s$-channel unitarity is discussed in sect.7.
Finally, in sect.8, we summarize the main results of the paper and discuss briefly possible
generalizations and applications of this approach.
\section{The model without diffusion}
We discuss in this section the $0$-dimensional model (without parton diffusion)
which posesses some typical features of the general case. This approximation
becomes adequate when it is possible to ignore parton diffusion, e.g.
in the regime of the uniform density distribution or in the case of interactions with
heavy nuclei.

Let us consider a system of partons with variable number of particles $N$,
which evolves in time $y$ due to the following elementary processes:
death of a parton, annihilation of two partons, splitting of any parton into two ones
and fusion of arbitrary parton pair into one parton.
Denote the probability of the $N$-parton state at the moment $y$ as $p_\N(y)$,
and the probabilities of death, annihilation, splitting and fusion of partons for the unit of time
as $m_1$, $m_2$, $\lambda$ and $\nu$, respectively.

The evolution equations have the form:
\begin{align}\label{kin0}
&{dp_\N\over dy} = - (\lambda +m_1)N\,p_\N - (\nu+m_2) N(N-1)\,p_\N +
\lambda (N-1) \,p_{\scss N-1} \nn \\
&~~ + m_1 (N+1)\, p_{\scss N+1} + \nu (N+1)N\, p_{\scss N+1}
+ m_2 (N+2)(N+1)p_{\scss N+2} ~, ~ (N\ge 2) ~, \nn \\
&{dp_0\over dy} = ~m_1 \, p_1 + 2 m_2\, p_2~, \nn \\
&{dp_1\over dy} = - (\lambda +m_1)\, p_1 + 2(m_1 +\nu)\, p_2 +6 m_2\, p_3 ~.
\end{align}
The initial condition for eqs (\ref{kin0}) is given by the parton distribution at the moment
$y=0$, and the conservation of the total probability $p(y)=\sum_{0}^{\infty}p_\N (y)$
follows from (\ref{kin0}):
\begin{align}
\label{ptot}
{dp\over dy} = 0 ~.
\end{align}

The equivalent set of equations can be written down for the factorial moments $\mu_s$,
defined as expectations of quantities $(N)_s\equiv \mbox{N!/(N-s)!}$:
\begin{align}
\mu_s = \sum_{N=0}^{\infty}(N)_s p_\N \equiv
\sum_{N=0}^{\infty} {N!\over (N-s)!}p_\N ~,
\qquad (s=0,1,2,\dots) ~.
\end{align}
After multiplying the both sides of the eqs (\ref{kin0}) to $(N)_s$ and expressing the
products appearing in the right-hand sides in terms of $(N)_s$, $(N)_{s-1}$, \dots,
we come to the set of equations for the moments:
\begin{align}
{d\mu_s \over dy} =& ~\lambda s(s-1) \mu_{s-1} + (\lambda - m_1) s \mu_s \nn\\
&
- (\nu + m_2) s(s-1) \mu_s - (\nu +2 m_2) s \mu_{s+1} ~.
\end{align}
In particular, the equation (\ref{ptot}) for $p(y)\equiv \mu_0(y)$ is reproduced
at $s=0$.

It is convenient to define the generating function $G(w;y)$, which is at the same time
the exponential generating function for moments:
\begin{align}\label{G0}
G(w;y) = \sum_{N=0}^{\infty} p_N (y)w^N
= \sum_{s=0}^{\infty} \mu_s (y) {(w-1)^s \over s!} ~.
\end{align}
It satisfies the partial differential equation:
\begin{align}\label{gen_0}
\frac{\d G} {\d y} = (1-w)(m_1-\lambda w) \frac{\d G} {\d w}
+(1-w)[m_2+(m_2+\nu) w]\frac{\d^2 G} {\d w^2}
\end{align}
with initial condition
\begin{align}\label{init}
G(w;0) \equiv G_0 (w) = \sum_{N=0}^{\infty} p_\N (0) w^N ~.
\end{align}
The boundary condition at $w=1$ is maintained automatically:
\begin{align}\label{bound1}
G(1;y) = 1 ~.
\end{align}
In case $m_1 = m_2 = 0$ it holds at $w=0$:
\begin{align}\label{bound0}
G(0;y) = G_0 (0) ~.
\end{align}

Let us consider some special cases:\\

\noindent
{\it (a) The absence of parton fusion and annihilation ($\nu = 0$, $m_2=0$)} \\

In this case Eq.(\ref{G0}) becomes of 1st order and
after the change of variables $w\rightarrow \gamma$, where
\begin{align}
d\gamma = {dw \over \lambda w^2 - (\lambda + m_1) w + m_1} ~,~~~
w(\gamma)={1 - (m_1 /\lambda)\exp[(\lambda - m_1)\gamma] \over
1-\exp[(\lambda - m_1)\gamma]} ~,
\end{align}
one obtains
\begin{align}
& {\d \hat{G}(\gamma ;y) \over \d y} = {\d \hat{G}(\gamma;y) \over \d\gamma} ~, \\
& \hat{G}(\gamma;0)  =  \hat{G}_0 (\gamma) ~, \qquad  \hat{G}(-\infty;y) = 1 ~,
\end{align}
with notation
$\hat{G} (\gamma ;y) \equiv G(w(\gamma);y), \hat{G}_0 (\gamma) \equiv G_0(w(\gamma))$.
Its solution is a function of the sum of the variables $\gamma+y$ satisfying the
initial condition (\ref{init}),
\begin{align}
\hat{G} (\gamma;y) = \hat{G} _0 (\gamma+y) ~,
\end{align}
or, coming back to the variable $w$,
\begin{align}\label{G0_1}
G(y,w) = \sum_{N} W(y)^N p_N(0) = G_0 (W(y)) ~,
\end{align}
where
\begin{align}\label{G0_2}
& W(y) = {1-\eta \,e^{\Delta (\gamma + y)}
\over 1- e^{\Delta (\gamma + y)}}
= {w-\eta -\eta (w-1) \,e^{\Delta y} \over w-\eta -(w-1) \,e^{\Delta y}} ~, \nn\\[3mm]
&
\Delta = \lambda -m_1 ~, \qquad \eta = m_1 / \lambda ~.
\end{align}
This distribution differs significantly from the Poisson distribution being characterized
by a presence of strong correlations.

If there is $N_0$ particles in the initial state, then
\begin{align}
G(w;y)=\left[ W(y)\right]^{N_0} ~,
\end{align}
The mean value of particles increases exponentially with energy in this case:
\begin{align}
\mu_1 (y) = N_0 e^{\Delta y} ~,
\end{align}
and the second moment behaves as follows:
\begin{align}
\mu_2 (y) = N_0 \left(N_0 +{1+\eta \over 1-\eta }\right) e^{2 \Delta y} - {2 N_0 \over 1-\eta } e^{\Delta y} ~.
\end{align}

In case when $m_1 \ne 0$, the state without partons ($N=0$) is so called
{\it absorbing state}. The probability $p_0(y)$ can only increase:
\begin{align}
p_0(y) = {\eta (e^{\Delta y}-1)\over e^{\Delta y} -\eta }
\end{align}
If the process of parton death dominates ($\Delta = \lambda - m_1 <0$), then
$p_0(y)\rightarrow 1$ at $y\rightarrow\infty$. If $\Delta > 0$, this probability
goes to the constant $\eta=m_1 / \lambda$.

In the case of $m_1 =0$ the probability $p_0(y)\equiv 0$, and the probability distribution
can be written as
\begin{align}
p_N (y) = \binom{\scs N-1}{\scs N-N_0}
e^{-\N_0 \Delta y}\left(  1- e^{-\Delta y}\right)^{\N-\N_0} ~.
\end{align}

\noindent
{\it (b) The asymptotic in the fusion presence ($\nu \ne 0, ~y\rightarrow \infty$)} \\

When $y\rightarrow\infty$ the state of the system approaches to the stationary state
$G_{\infty}(w)=\lim_{y\rightarrow\infty} G(w;y)$ of eq.(\ref{gen_0}), which can be found
from the equation for $\phi (w) \equiv dG_{\infty}/dw$:
\begin{align}\label{stat_0}
[m_2+w(m_2+\nu)] {d \phi  \over d w} + ( m_1 - \lambda w )\phi = 0 ~.
\end{align}
Solutions of this equation have the form
\begin{align}\label{gen}
\phi = C e^{a w} (w+c)^{-b} ~,
\end{align}
where $a = \lambda/(m_2+\nu)$ , $b=(m_1 m_2+m_1\nu+m_2\lambda)/(m_2+\nu)^2$,
$c=m_2/(m_2+\nu)$, and $C$ is determined from the initial condition.
This gives
\begin{align}\label{G_as}
G_{\infty}(w)&= 1+\int_1^w \phi(x)dx = 1+ C[F(w) - F(1)] ~,
\end{align}
with
\begin{align}\label{F}
F(w)=\frac{e^{a(w+c)}}{a(w+c)^b}-(-a)^{b-1}b\Gamma(-b,-a(w+c))~,
\end{align}
where $\Gamma$ is the standard Gamma function of two arguments.

If $m_1=m_2=0$ we get
\begin{align}\label{as_0}
G_{\infty}(w)=1-B+B e^{a (w-1)} ~,
\end{align}
with the constant $B$ determined by the initial condition.

Thus, the state of the system becomes asymptotically Poisson-like:
\begin{align}\label{asymp}
& \mu_0 \xrightarrow[y\to\infty]{} 1 ~,\qquad \mu_s \xrightarrow[y\to\infty]{} B a^{s} ~, \quad s \ge 1  ~, \nn \\
& p_0(y) = 1-B+B e^{-a} ~, \qquad
p_N \xrightarrow[y\to\infty]{} B {a^\N \over N!} e^{-a} ~, \quad N\ge 1 ~~~.
\end{align}
Though the moments $\mu_s$ depend on the subscript $s$ in power way
the relations between moments reveal the correlation connected to independence of $p_0$
from other states:
$$
\mu_1 (\infty)= Ba ~, \qquad \mu_s(\infty) = Ba^s \ne \left[\mu_1(\infty)\right]^s ~.
$$
At $B=1$ one gets the purely Poisson distribution and all correlations vanish.

The value of the constant $B$ which determines the asymptotical behaviour
depends on a choice of the initial condition. Due to (\ref{bound0})
\begin{align}
B={1-G_0 (0) \over 1- e^{-a}} ~.
\end{align}
Note that $G_0 (0)=p_0(0)$ and if $p_0(0)=p_0(y)=0$, then $B=(1- e^{-a})^{-1}$.

If $m_1$ (or $m_2$) differs from zero the state with $N=0$ is the {\em absorbing} one.
The death or annihilation processes dominate asymptotically in this case and, eventually,
the system goes to the state without particles:
\begin{align}
p_\N \xrightarrow[y\to\infty]{}\delta_{\N0} ~,
\qquad G_{\infty}(w) = 1  ~.
\end{align}
This conclusion can be done from direct analysis of the kinetic equations (\ref{kin0}),
after putting to zero the derivatives in the left-hand sides.

Note that if $a = \lambda/\nu \gg 1$, the average number of particles in the system
is large and its state can be described preasymptotically in thermodynamical approximation
by eq.(\ref{asymp}), though eventually at very large $y$ it goes to the absorbing state.
\section{Account for spacial diffusion of partons}

In this section we take into account a possibility of random walk of partons
in the transverse plane in the course of evolution. We assume that a transverse coordinate
of a parton, $\b_i$, gets for an infinitely small interval $dy$ an (isotropic) increment
$d\b_i$ such that $(d\b_i)^2=D\, dy$. Note that this assumption on the character of the random
walk of partons results in the linear form of the pomeron trajectory. It is
a hypothesis only and other dynamical models for parton diffusion are conceivable
\footnote{~One can consider a discrete model of parton diffusion by granulation the transverse
space into small cells. In this case the random walk is included to the kinetic scheme
on equal footing with splitting and fusion. The size of a cell plays the role of a cut-off
for the model.}.

Let us formulate now dynamical equations for the parton system. Denote the probability
for the system to be in the state with $N$ partons at the moment $y$ as $p_\N(y)$,
and the parton distribution in the impact plane as $\rho_\N (y;\b_1,\dots,\b_\N)$.
To simplify the notations we shall omit the argument $y$ often, and denote the set of $N$
transverse coordinates as $\B_\N\equiv \{\b_1,\dots,\b_\N\}$.
All partons are supposed to be of the same type
\footnote{~If the internal quantum numbers of partons are introduced then relations between
the constants of the RFT corresponding to given parton model, particularly between the intercept
and the triple pomeron coupling, will be changed.},
and $\rho_\N$ is a symmetric function of coordinates normalized as
\begin{align}\label{norm_rho}
{1\over N!}\int \!d\B_\N \rho_\N (y;\B_\N) = p_\N (y) ~,
\end{align}
where $d\B_\N=d\b_1\ldots d\b_\N$, and $ p_\N (y)$ is the probability to have the $N$-parton
state at the moment $y$.

The following changes of the state of the system for an interval $dy$ are possible:
change of position of any parton to $(d\b_i)^2=D\, dy$; {\em death} of a parton with the probability
$m_1 \, dy$; {\em splitting} of one parton into two ones with same coordinates with the probability
$\lambda \, dy$, and, for two partons sufficiently close one to another, processes of their
{\em annihilation} with vanishing both partons and their {\em fusion} into one parton.
Assuming that the parton size is small enough (compared to interparton distances)
the annihilation and fusion probability of partons at points $\b_k$ and $\b_l$ can be described as
$m_2\, \delta (\b_k-\b_l)$ and $\nu\, \delta (\b_k-\b_l)$ correspondingly
\footnote{~As it was mentioned, one can consider parton coordinates as discrete quantities
in order to regularize divergences of the model. This results in the correspondingly smoothed
 $\delta$ functions.}.
Let us stress that the constants $m_2$ and $\nu$ are dimensional in contrast to the situation of sect.2.

Thus, the evolution equations for the densities $\rho_\N$ have the following form:
\begin{align}\label{ev_rho}
&{d\rho_\N (y;\B_\N)\over dy} =
\,D \,\nabla_{\N}^2 \,\rho_\N (y;\B_\N) - (m_1+\lambda) N \rho_\N (y;\B_\N) \nn \\
&\phantom{d}
+ m_1 \int\! db_{\scss N+1} \rho_{\scss N+1}(y;\B_\N,\b_{\scss N+1})
+\lambda \sum_{k,l=1}^{\scss N\geqslant 2} \rho_{\scss N-1} (y;\B_{\N}^{(l)}) \delta (\b_k-\b_l) \nn \\
&\phantom{d}
- (m_2 +\nu ) \sum_{k,l=1}^{\scss N\geqslant 2} \rho_\N (y;\B_\N) \delta (\b_k-\b_l)
+ m_2 \int\! d\b_{\scss N+1} \rho_{\scss N+2}(y;\B_\N,\b_{\scss N+1},\b_{\scss N+1})
\nn \\
&\phantom{d}
+ \nu \int\! d\b_{\scss N+1} \rho_{\scss N+1}(y;\B_\N,\b_{\scss N+1})
\sum_{k=1}^{\N} \delta (\b_k-\b_{\scss N+1}) ~,
\end{align}
where $\nabla_{\N}^2 =
\sum_{i=1}^{N} (\partial_{\alpha}^2 \rho_\N /\partial z_{i\alpha}^2), \alpha=1,2$, and
$\B_{\N}^{(l)}$ means the set $\B_\N$, from which the coordinate $\b_l$ is removed
($\B_{\N}^{(l)}=\B_\N\setminus\b_l$).

One can also define the Fourier transformed distributions
\begin{align}
\sigma_\N (y;\K_\N) = \int\! d\B_\N \rho_\N (y;\B_\N)
e^{i\k_1\b_1+\cdots +i\k_\N\b_\N} ~
\end{align}
where $\K_\N=\{ \k_1,\ldots,\k_\N\}$, with normalization
\begin{align}\label{norm_sig}
\sigma_\N (y;0,\ldots,0) = N! \, p_\N (y) ~.
\end{align}
In the momentum representation the evolution equations take the form:
\begin{align}\label{ev_sig}
&{d\sigma_\N(y;\K_\N)\over dy} =
- D\; \left( \sum_{i=1}^\N \k_i^2\right)\; \sigma_\N (y;\K_\N) -
(m_1 + \lambda) N  \sigma_\N (y;\K_\N) \nn \\
&\phantom{d\sigma_\N(y}
+ m_1 \, \sigma_{\scss N+1} (y;\K_\N,\bmath{0}) +
\lambda \sum_{k,l=1}^{\scss N\geqslant 2}\sigma_{\scss N-1} (y;\K_{\N}^{(kl)}, \k_k +\k_l) \nn \\
&\phantom{d\sigma_\N(y}
- (m_2 + \nu)\! \int\! {d\q \over (2\pi)^2} \sum_{k,l=1}^{\scss N\geqslant  2}
\sigma_{\N}(y;\K_{\N}^{(kl)},\k_k+\q,\k_l-\q)\nn \\
&\phantom{d\sigma_\N(y}
+ m_2 \int\! {d\q \over (2\pi)^2}
\sigma_{\scss N+2}(y;\K_{\N},\q,-\q)
\nn \\
&\phantom{d\sigma_\N(y}
+ \nu \int\! {d\q \over (2\pi)^2} \sum_{k=1}^\N
\sigma_{\scss N+1}(y;\K_{\N}^{(k)},\k_k+\q,-\q) ~,
\end{align}
where $\K_{\N}^{(k)}=\K_\N\setminus \k_k$ and $\K_{\N}^{(kl)}=\K_\N\setminus\{\k_k,\k_l\}$
are the sets of momenta $\K_\N$ with the momentum $\k_k$ or, respectively, the momenta $\k_k$
and $\k_l$, removed.

It is convenient, in the spirit of statistical mechanics, to introduce instead of distributions
$\rho_\N (y,\B_\N)$ a set of the multiparticle distributions
$f_s^{\scss (N)} (y; z_1,\ldots z_s), s=1,\dots,N$,
which correspond to fixation of the coordinates of $s$ partons and integration over
coordinates of all rest:
\begin{align}\label{def_f}
f_s^{\scss (N)}(y;\Z_s)={1\over N!} \int\! d\B_{\N}\;\rho_\N(y;\B_\N) A_s^{\scss (N)}(\B_N | \Z_s) ~,
\end{align}
where, as above, $\Z_s=\{ \z_1,\ldots,\z_s \}$. Here the function
\begin{align}
A_s^{\scss (N)}(\B_N | \Z_s) = \sum_{\II_s^{\scss (N)}}
\delta (\z_1 -\b_{i_1})\delta (\z_2 -\b_{i_2})\dots \delta (\z_s -\b_{i_s}) ~,
\end{align}
where summation is taken over all sets $\II_s^{\scss (N)}$ of noncoincident indices
$i_1, i_2 \dots i_s$ from the set $\{1,2,\dots,N\}$, is a sum of $\delta$-function terms
which fix coordinates of $s$ partons from $N$ ones.

The total multiparticle distributions are defined by summation over states with different
number of partons:
\begin{align}
 f_s(y;\Z_s)=\sum_N  f_s^{\scss (N)}(y;\Z_s) ~.
\end{align}
The normalization of the functions $f_s^{\scss (N)}(y;\Z_s)$ follows from eq.(\ref{norm_rho}):
\begin{align}
\int\! d\Z_s f_s^{\scss (N)}(y;\Z_s) = (N)_s p_\N ~,
\end{align}
where, as above, $(N)_s=N!/(N-s)!~$. Correspondingly, the distribution $ f_s(y;\Z_s)$
is normalized to the mean factorial moment:
\begin{align}
\int\! d\Z_s f_s (y;\Z_s) = <(N)_s>\equiv \mu_s(y) ~.
\end{align}
The functions $f_s^{\scss (N)}(y;\Z_s)$ are related by the reduction equation:
\begin{align}
f_{s-r}^{\scss (N)}(y;\Z_{s-r}) = {(N-s)!\over (N-r)!} \int\! dz_{r+1}\dots dz_s
f_s^{\scss (N)}(y;\Z_s) ~.
\end{align}

Quite similarly, one can introduce the momentum multiparticle distributions
$g_s^{\scss (N)}(y;\Q_s)$, connected to the functions $ f_s^{\scss (N)}$
with Fourier transform:
\begin{align}
& g_s^{\scss (N)}(y;\Q_s)=
\int\!  d\Z_s e^{i\q_1 \z_1}\dots  e^{i\q_s \z_s}  f_s^{\scss (N)}(y;\Z_s) ~, \nn\\
& g_s(y;\Q_s)=\sum_N  g_s^{\scss (N)}(y;\Q_s) ~.
\end{align}
These functions are obtained from the distributions $\sigma_\N (y;\K_\N)$, if to put
$s$ momenta equaled to $q_1,\dots,q_s$, and all the rest --- to zeros:
\begin{align}
 g_s^{\scss (N)}(y;\Q_s)={1\over (N-s)!}\,\sigma_\N (q_1,\dots,q_s,0,\dots,0) ~.
\end{align}
The reduction relation has the form:
\begin{align}\label{red_g}
 g_{s-r}^{\scss (N)}(y;\Q_{s-r}) = {(N-s)!\over (N-r)!}\,
g_s^{\scss (N)}(y;\Q_{s-r},\underbrace{0,\dots,0}_{\text{$r$ times}}) ~,
\end{align}
and normalization looks as
\begin{align}
& g_s^{\scss (N)}(y;0,\dots,0) =  (N)_s p_\N ~, \nn\\
& g_s(y;0,\dots,0) = \mu_s(y) ~.
\end{align}

The simplest way to get the kinetic equations for multiparticle distributions
is to reduce eq.(\ref{ev_sig}) by means of eq.(\ref{red_g}) from the function
$g_\N^{\scss (N)}\equiv \sigma_\N$ to the function $g_s^{\scss (N)}$. This gives
\begin{align}
&{d\over dy} g_s^{\scss (N)}(y;\Q_s) =
- D\left(  \sum_{a=1}^s \q_a^2\right) g_s^{\scss (N)}(y;\Q_s)
- (m_1 +\lambda)N g_s^{\scss (N)}(y;\Q_s) \nn\\
&\phantom{{d\over dy}}
+ m_1 (N-s+1) g_s^{\scss (N+1)}(y;\Q_s)
+ \lambda (N+s-1) g_s^{\scss (N-1)}(y;\Q_s)
\nn\\
&\phantom{{d\over dy}}
+\lambda \sum_{k,l=1}^{s\geqslant  2} g_{s-1}^{\scss (N-1)}(y;\Q_s^{(k,l)},\q_k+\q_l)
\nn\\
&\phantom{{d\over dy}}- (m_2+\nu)\! \int\! {d\q \over (2\pi)^2} \sum_{k,l=1}^{s\geqslant 2}
g_s^{\scss (N)}(y;\Q_s^{(k,l)},\q_k+\q,\q_l-\q) \nn\\
&\phantom{{d\over dy}}
- 2 (m_2+\nu )\! \int\! {d\q \over (2\pi)^2} \sum_{k=1}^s
g_{s+1}^{\scss (N)}(y;\Q_s^{(k)},\q_k+\q,-\q) \nn\\
&\phantom{{d\over dy}}
- (m_2 + \nu )\! \int\! {d\q \over (2\pi)^2} g_{s+2}^{\scss (N)}(y;\Q_{s},\q,-\q)
+ m_2 \! \int\! {d\q \over (2\pi)^2} g_{s+2}^{\scss (N+2)}(y;\Q_{s},\q,-\q) \nn\\
&\phantom{{d\over dy}}
+ \nu\! \int\! {d\q \over (2\pi)^2} \sum_{k=1}^s
g_{s+1}^{\scss (N+1)}(y;\Q_s^{(k)},\q_k+\q,-\q)
\nn\\
&\phantom{{d\over dy}}
 +  \nu\! \int\! {d\q \over (2\pi)^2} g_{s+2}^{\scss (N+1)}(y;\Q_{s},\q,-\q) ~.
\end{align}
Note here, that when doing the reduction one has to consider separately the cases when
summation variables $i,k$ are less and greater than $s$, and this increases a number
of terms with coefficients $\lambda$ and $\nu$. The combinatorial coefficients appear
in the equation due to this consideration (the common factor $(N-s)!$ is omitted
in the equation).

Part of terms cancel after summation over $N$ (particularly terms containing $g_{s+2}$)
and one gets
\begin{align}\label{ev_g}
&{d\over dy} g_s (y;\Q_s ) =
- D \left(  \sum_{a=1}^s \q_a^2\right) g_s(y;\Q_s)
+ (\lambda - m_1 ) s g_s (y;\Q_s) \nn\\
&\phantom{{d\over dy}}
\p+\lambda\! \sum_{k,l=1}^{s\geqslant  2} g_{s-1} (y;\Q_s^{(k,l)},\q_k+\q_l)
\m-  \nu\! \int\! {d\q \over (2\pi)^2} \sum_{k,l=1}^{s\geqslant 2}
g_s (y;\Q_s^{(k,l)},\q_k\p+\q,\q_l\m-\q) \nn\\
&\phantom{{d\over dy}}
\m- (2m_2+\nu)\! \int\! {d\q \over (2\pi)^2} \sum_{k=1}^s
g_{s+1} (y;\Q_s^{(k)},\q_k+\q,-\q) ~, ~~ (s=1,2,\ldots) ~.
\end{align}

The equation for $f_s (y;\Z_s)$ can be obtained both by Fourier transform
of (\ref{ev_g}) or directly from the evolution equation (\ref{ev_rho}) for $\rho_N$
using the definition (\ref{def_f}) (it is necessary again to account for accurate
combinatorics for cases $i,k \le s$ and $i,k > s$). As a result one gets the following
equations for $f_s^{\scss (N)}$:
\begin{align}
&{d\over dy}  f_s^{\scss (N)} (y; \Z_s) =
\, D \,\vec{\nabla}_s^2 f_s^{\scss (N)} (y; \Z_s)
- (m_1+\lambda) N  f_s^{\scss (N)} (y; \Z_s)
\nn \\ &\phantom{{d\over dy}}
+ m_1 (N-s+1) f_s^{\scss (N+1)} (y; \Z_s) +
\lambda (N-1+s) f_s^{\scss (N-1)} (y; \Z_s)
\nn \\ &\phantom{{d\over dy}}
+ \lambda\! \sum_{k,l=1}^{s\geqslant  2} f_{s-1}^{\scss N-1} (y;\Z_s^{(l)}) \delta (\z_k -\z_l)
- (m_2 +\nu )\! \sum_{k,l=1}^{s\geqslant  2} f_s^{\scss (N)} (y; \Z_s) \delta (\z_k -\z_l)
\nn \\ &\phantom{{d\over dy}}
- 2 (m_2 +\nu )\!  \sum_{k=1}^s f_{s+1}^{\scss (N)} (y; \Z_s^{(k)},\z_k,\z_k)
- (m_2 +\nu )\! \int\! d\u f_{s+2}^{\scss (N)} (y; \Z_s,\u,\u)
\nn\\ &\phantom{{d\over dy}}
+ m_2 \int\! d\u f_{s+2}^{\scss (N+2)} (y; \Z_s,\u,\u)
+ \nu\!  \sum_{k=1}^s f_{s+1}^{\scss (N+1)} (y; \Z_s^{(k)},\z_k,\z_k)
\nn \\ &\phantom{{d\over dy}}
+ \nu\!  \int\! d\u f_{s+2}^{\scss (N+1)} (y; \Z_s,\u,\u)  ~,
\end{align}
and the set of equations for the total distribution functions $f_s (y;\Z_s)$:
\begin{align}\label{ev_f}
&{d\over dy}  f_s (y; \Z_s) \r=
\, D\, \vec{\nabla}_s^2 f_s (y; \Z_s) \p+ (\lambda - m_1) s  f_s (y; \Z_s) +
\nn\\ &\phantom{{d\over dy}}
\p+ \lambda\! \sum_{k,l=1}^{s\geqslant  2} f_{s-1} (y;\Z_s^{(l)}) \delta (\z_k -\z_l)
\m- \nu\! \sum_{k,l=1}^{s\geqslant  2} f_s (y; \Z_s) \delta (\z_k - \z_l) -
\nn\\ &\phantom{{d\over dy}}
\m- (2 m_2 +\nu )\!  \sum_{k=1}^s f_{s+1} (y; \Z_s^{(k)},\z_k,\z_k) ~, \qquad (s=1,2,\ldots) ~.
\end{align}

Thus, we came to equations for the sets of the functions $f_s (y;\Z_s)$ or $g_s (y;\Q_s)$
which contain only distributions averaged over states with various numbers of partons.
\section{Equivalence of the parton model to reggeon field theory}

It will be shown in this section that eqs (\ref{ev_g}), (\ref{ev_f}) are equivalent to equations
for reggeon vertices of the reggeon field theory with a particular set of coupling constants
--- the reggeon intercept $\Delta=\lambda -m_1$, the trajectory slope $\alpha' = D$,
the constant of splitting of one pomeron into two ones $\lambda$,
the constant of fusion of two pomerons into one $2 m_2 +\nu$ and
the constant of pomeron scattering equaled to $\nu$.

The simplest way to see this, is to make the Mellin transform of the function $g_s(y)$:
\begin{align}
G_s(\omega;\Q_s)=\int_{0}^{\infty}\!\! dy e^{-\omega y} g_s (y;\Q_s)~, ~
g_s(y,\Q_s) = \frac{1}{2\pi i}
\!\int_{\uparrow}\! d\omega e^{\omega y} G_s(\omega;\Q_s) ~,
\end{align}
where the integration in second formula is carried out in the complex plane $\omega$
along the infinite contour parallel to the imaginary axis.

One gets the following set of equations for the functions $G_s (\omega )$
\begin{align}\label{omega}
\omega & G_s(\omega ;\Q_s) =  D\left( \sum_{a=1}^{s} q_a^2 \right) G_s(\omega;\Q_s) +
(\lambda - m_1) sG_s(\omega;\Q_s) \nn\\
&+ \lambda \sum_{k,l=1}^{s} G_{s-1}(\omega;\Q_s^{(kl)},\q_k\p+\q_l)
- \nu \int\! \frac{d\q}{(2\pi)^2}\sum_{k,l=1}^{s}
 G_s(\omega;\Q_s^{(kl)},\q_k\p+\q,\q_l\m-\q)  \nn\\
&- (2 m_2+\nu) \int\! \frac{d\q}{(2\pi)^2}\sum_{k=1}^{s} G_{s+1}(\omega;\Q_s^{(k)},\q_k\p+\q,\m-\q) ~.
\end{align}
Let the initial condition for (\ref{ev_g}) be
\begin{align}
g_s(y=0;\Q_s) = \delta_{s s_0} ~.
\end{align}
Then $g_s(y;\Q_s)$ (respectively $G_s (\omega ;\Q_s)$ in $\omega$ representation) have
meaning of vertices of transition of $s_0$ reggeons into $s$ reggeons. The function of
free propagation (diffusion without splitting, annihilation  and fusion) of $s$ reggeons
in the $\omega$ representation has the form
\begin{align}\label{D_s0}
D_s^{(0)}(\omega ;\Q_s) = \left( \omega - D \sum_{a=1}^{s} q_a^2 \right)^{-1} ~,
\end{align}
and the renormalized propagator can be defined as
\begin{align}\label{D_s}
D_s (\omega ;\Q_s) = \left( \omega - s (\lambda - m_1) - D \sum_{a=1}^{s} q_a^2 \right)^{-1} ~.
\end{align}
If define the reduced vertices
\begin{align}
\Gamma_{s}^{s_0}(\omega ;\Q_s) = \left[ D_s (\omega ;\Q_s)\right]^{-1} G_s (\omega ;\Q_s) ~,
\end{align}
the equations (\ref{omega}) take the form
\begin{align}\label{eqs_RFT}
\Gamma_{s}^{s_0}&(\omega ;\Q_s) =
\lambda \sum_{k,l=1}^{s} \Gamma_{s-1}^{s_0}(\omega ;\Q_s^{(kl)},\q_k+\q_l) D_{s-1}(\omega ;\Q_s) - \nn\\
- & \nu \int\! \frac{d\q}{(2\pi)^2}\sum_{k,l=1}^{s}
\Gamma_{s}^{s_0}(\omega;\Q_s^{(kl)},\q_k+\q,\q_l-\q) D_{s}(\omega ;\Q_s) - \nn\\
- & (2 m_2+\nu) \int\! \frac{d\q}{(2\pi)^2}\sum_{k=1}^{s}
\Gamma_{s+1}^{s_0}(\omega;\Q_s^{(k)},\q_k+\q,-\q) D_{s+1}(\omega ;\Q_s) ~.
\end{align}
which corresponds to the rules of reggeon diagram technique \cite{gribov67} and is described
by diagrams of fig.{\it 1}.
\begin{figure}[h]
\begin{center}
\begin{picture}(300,60)(10,30)
\GOval(10,70)(10,18)(0){1}
\Photon(18,95)(18,79){2}{3}
\Photon(10,95)(10,80){2}{3}
\Photon(2,95)(2,79){2}{3}
\Photon(15,40)(14,60){2}{3}
\Photon(5,40)(6,60){2}{3}
\Photon(-4,40)(-2,62.5){2}{3}
\Photon(24,40)(22,62.5){2}{3}
\Text(10,70)[cc]{$\scriptstyle\Gamma_{s}^{s_{\scriptscriptstyle{0}}}$}
\Text(50,71)[cc]{{\large $= \;\lambda$}}
\GOval(90,70)(10,18)(0){1}
\Photon(98,95)(98,79){2}{3}
\Photon(90,95)(90,80){2}{3}
\Photon(82,95)(82,79){2}{3}
\Photon(76,40)(78,62.5){2}{3}
\Photon(104,40)(102,62.5){2}{3}
\Vertex(90,50){1.5}
\Photon(90,51)(90,60){1.2}{2}
\Photon(90,49)(96,40){1.2}{2}
\Photon(90,49)(84,40){-1.2}{2}
\DashLine(68,57)(108,57){3}
\Text(120,57)[cc]{$\scriptstyle D_{s-1}$}
\Text(92,70)[cc]{$\scriptstyle\Gamma_{s-1}^{s_{\scriptscriptstyle{0}}}$}
\Text(130,70)[cc]{{\large $- \;\nu$}}
\GOval(170,70)(10,18)(0){1}
\Photon(178,95)(178,79){2}{3}
\Photon(170,95)(170,80){2}{3}
\Photon(162,95)(162,79){2}{3}
\Photon(156,40)(158,62.5){2}{3}
\Photon(184,40)(182,62.5){2}{3}
\Vertex(170,50){1.5}
\Photon(170,50)(164,60.5){1.2}{2}
\Photon(170,50)(176,60.5){-1.2}{2}
\Photon(170,50)(176,40){1.2}{2}
\Photon(170,50)(164,40){-1.2}{2}
\DashLine(150,57)(190,57){3}
\Text(198,57)[cc]{$\scriptstyle D_{s}$}
\Text(170,70)[cc]{$\scriptstyle\Gamma_{s}^{s_{\scriptscriptstyle{0}}}$}
\Text(230,70)[cc]{{\large $-(2m_2\p+\nu)$}}
\GOval(290,70)(10,18)(0){1}
\Photon(298,95)(298,79){2}{3}
\Photon(290,95)(290,80){2}{3}
\Photon(282,95)(282,79){2}{3}
\Photon(276,40)(278,62.5){2}{3}
\Photon(304,40)(302,62.5){2}{3}
\Vertex(290,50){1.5}
\Photon(290,50)(296,60.5){-1.2}{2}
\Photon(290,50)(284,60.5){1.2}{2}
\Photon(290,50)(290,40){1.2}{2}
\DashLine(268,57)(308,57){3}
\Text(320,57)[cc]{$\scriptstyle D_{s+1}$}
\Text(290,70)[cc]{$\scriptstyle\Gamma_{s+1}^{s_{\scriptscriptstyle{0}}}$}
\end{picture}
\end{center}
\caption{Diagrammatic representation of equations for the inclusive parton distributions
which reproduces diagrams of RFT}
\end{figure}
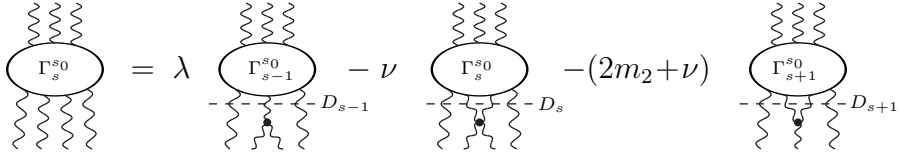

Another correspondence of eqs (\ref{eqs_RFT}) with reggeon theory is their treatment
as equations of ``inelastic'' reggeon theory induced by Abramovsky-Gribov-Kancheli cutting
rules~\cite{AGK} (see details in ref.~\cite{Boreskov89}).
Suppose that the contribution of some reggeon graph ${\G}$
to the scattering amplitude is schematically represented as
$-i\hat{T}^{(\G)}(\{- D_k\})$, i.e. depends on the set of reggeon propagators $\{D(y_k,\q_k)\}$
integrated with required weights. Its contribution to the total cross section is
$\sigma^{tot(\G)}/2=\Im[-i\hat{T}^{(\G)}(\{- D_k\})]$. Then (see ref.~\cite{Boreskov89})
the inelastic discontinuity of this diagram, i.e.
its contribution to the {\em inelastic} cross section, can be written as
$\sigma^{inel(\G)}=-\hat{T}^{\G}(\{-2\Im (iD_k)\})$, that is it also can be calculated by means
of reggeon theory but with {\em cut} propagators.
Contributions of the diagram ${\G}$ to generating functions
for different inelastic characteristics also can be determined from this theory~\cite{Boreskov89}
by changing arguments of operator $\hat{T}^{\G}$.

Thus, evolution of the parton distributions in rapidity is described
by diagrams of reggeon field theory. The pomerons do not come out in this
description as some individual propagating objects. The evolution in rapidity of the
$s$-pomeron state corresponds to the developing in time the $s$-particle {\em inclusive
distribution} of partons (i.e. when momenta of $s$ partons are fixed and momenta
of all others are arbitrary).
\section{Interaction of colliding hadrons}

In previous sections the parton model was formulated which dynamics is equivalent
to reggeon field theory. Parton distributions at different $y$ given by this model
can be associated with the stationary structure of parton distributions in rapidity
for a fast moving hadron (i.e. with its Fock wave function).

We define in this section the ``interaction operator'' of two parton systems,
which allows to calculate the amplitude for interactions of two fast
colliding hadrons. The form of this operator has to meet the requirement
of correspondence with calculations according to the RFT rules.

Let two hadrons $h$ and $\tilde{h}$ move one toward another with rapidities
$y$ and $\y$ in some Lorentz frame at impact parameter $\b$.
The essential feature required of the interaction operator is that Lorentz invariance
of the scattering amplitude, that is, its independence from a choice
of particular Lorentz frame, should be maintained.
In other words, the amplitude should be a function of the total rapidity $Y=y+\y$ only
\footnote
{~Though the hadrons move in the opposite directions, we define both $y$ and $\y$ as positive.
So the total rapidity which is the rapidity value of one hadron in the rest frame of other
is written as a sum $Y=y+\y$.}
, but not of the individual values of $y$ and $\y$.

The hadron states are specified by sets of parton distributions
$\{f_s (y;\Z_s)\}$ and $\{\f_s (\y;\tilde{\Z}_s+\b)\}$,
where designation $\tilde{\Z}_s+\b$ means a shift of all coordinates $\tilde{\z}_a$
to the vector $\b$ in the impact plane.
Denote for brevity these sets as vectors in space of states $|\F(y;0)\rangle$
and $|\tilde{\F}(\y;\b)\rangle$. In the momentum representation we shall use notation
$|\G(y;0)\rangle$ and $|\tilde{\G}(\y;\b)\rangle$ for the sets $\{g_s (y;\Q_s)\}$ and
$\{\g_s (\y;\Q_s)\exp(i \b\sum_1^s \q_a)\}$.

Let the total cross section is determined as an integral of function $T(y,\y;b)$
over all impact parameter values:
\begin{align}\label{Tb}
\sigma^{(tot)}(y,\y) = 2\int d^2b\, T(y,\y;b) ~,
\end{align}

We introduce the ``interaction operator'' $\T$ by the following equation:
\begin{align}\label{T_f}
&T(y,\y;b) = \langle \tilde{\F}(\y;\b) | \T |\F(y;0)\rangle = \nn \\
&~~~ = \sum_{s=1}^{\infty} {(-1)^{s-1}\epsilon^s \over s!}
\int d{\tilde \Z}_s d\Z_s \f_s (\y;{\tilde \Z}_s) f_s (y;\Z_s)
\delta^{(s)} (\Z_s - {\tilde \Z}_s - \b) ~,
\end{align}
or, in momentum space,
\begin{align}\label{T_g}
T(y,\y;b) &= \langle \tilde{\G}(\y;\b) | \T |\G(y;0)\rangle = \nn \\
&=  \sum_{s=1}^{\infty} {(-1)^{s-1}\epsilon^s \over s!}
\int \frac{d\Q_s}{(2\pi)^{2s}}\, e^{i \b\sum_1^s \q_a} \g_s (\y;\Q_s) g_s (y;\Q_s)\, ~.
\end{align}
The parameter $\epsilon$ plays a role of the squared size of a parton. As we mentioned in sect.3
in order to regularize some singular expressions the function $\epsilon^s \delta^{(s)}$ should
be smeared out over this range.

Let us show that the definition (\ref{T_f}), (\ref{T_g}) of the operator $\T$ combined
with particular relation between constants provides the dependence of the
function $T(y,\y;b)$ only from the single variable $Y=y+\y$.
In the momentum representation the evolution equations (\ref{ev_g}) can be symbolically written as
\begin{align}
{d\over dy}|\G(y;\b)\rangle = \H |\G(y;\b)\rangle ~,
\end{align}
where the form of the operator $\H$ follows from eq.(\ref{ev_g}).
Its solution is
\begin{align}
|\G(y;\b)\rangle = \exp \left[ \H y  \right] |\G(0;\b)\rangle ~.
\end{align}
i.e. the evolution operator $\exp \left[ \H y  \right]$ has a meaning of the Lorentz boost
operator. Therefore
\begin{align}
T(y,\y;b) = \langle \G(0;b)| e^{\H^t \y} \T e^{\H y} |G(0;0)\rangle ~,
\end{align}
($\H^t$ is the transpose of $\H$).
If the relation
\begin{align}\label{prop}
\langle\chi|\H^t \T |\phi\rangle = \langle\chi| \T \H |\phi\rangle ~
\end{align}
holds for arbitrary states $|\phi>,|\chi>$, then the required property will be fulfilled:
\begin{align}
\langle\chi| e^{\H^t \y} \T e^{\H y} |\phi\rangle = \langle\chi| \T e^{\H (y+\y)} |\phi\rangle ~.
\end{align}
Let us prove the relation (\ref{prop}).
The first, second, and fourth terms of the matrix $\H$ (see eq.(\ref{ev_g})) are diagonal in the parton numbers
$s$, and the relation (\ref{prop}) is evident. The third and fifth
terms (denote them as $H_{12}$ and $H_{21}$) correspond to splitting and fusion of reggeons
and their matrix elements have to be conjugated to each other, i.e.
$\langle\chi| T H_{21} |\phi\rangle = \langle\chi| H_{12}^t\,T |\phi\rangle$ and vice versa.
Indeed, the matrix element of $T H_{21}$ is written as
\begin{align}
\langle\chi| T H_{21} |\phi\rangle =&
-(2 m_2 +\nu) \sum_{s=1}^{\infty} {(-1)^{s-1}\epsilon^s \over s!}
\int \frac{d\Q_s}{(2\pi)^{2s}}\, {d\q \over (2\pi)^2}\chi_s(\Q_{s})
\nn\\&
\times\sum_{k=1}^s \phi_{s+1} (\Q_s^{(k)},\q_k+\q, -\q)
= (2 m_2 +\nu) \sum_{s=1}^{\infty} {(-1)^{s}\epsilon^s \over (s-1)!}
\nn\\ &
\times\int \frac{d\Q_s}{(2\pi)^{2s}}\, {d\q \over (2\pi)^2}
\chi_s(\Q_{s}) \phi_{s+1} (\Q_s^{(k)},\q_k+\q, -\q) ~,
\end{align}
with regard that the sum over $k$ contains $s$ identical terms.
In the matrix element of $T H_{12}$ summation over parton number starts with $\s=2$.
Hence, after substitution $\s=s+1$  and taking into account that the sum over $k,l$ contains
$s(s+1)$ identical terms one gets
\begin{align}
 \langle\chi| H_{12}^t\,T |\phi\rangle =& \langle\phi| T H_{12} |\chi\rangle =
\lambda \sum_{\s=2}^{\infty} {(-1)^{\s-1}\epsilon^{\s} \over \s !}
\int \frac{d\Q_{\s}}{(2\pi)^{2\s}} \phi_{\s}(\Q_{\s})
\nn\\&
\times\sum_{k,l=1}^{\s} \chi_{\s-1} (\Q_{\s}^{(kl)},\q_k+\q_l) =
\lambda \,\epsilon  \sum_{s=1}^{\infty} {(-1)^{s}\epsilon^s \over (s-1)!}
\nn\\&
\times\int \frac{d\Q_{s+1}}{(2\pi)^{2(s+1)}}
\chi_{s} (\Q_{s-1},\q_s+\q_{s+1}) \phi_{s+1}(\Q_{s+1}) ~.
\end{align}
After change of integration variables the both matrix elements are the same provided
\begin{align}\label{relat}
2 m_2 +\nu = \lambda \,\epsilon ~.
\end{align}
In the case of the $0$-dimensional
model constants $\lambda$, $m_2$ and $\nu$ are dimensionless, the parameter $\epsilon$
in the definition of the interaction operator is not needed and the requirement of Lorentz
invariance gives the relation
\begin{align}\label{relat0}
2 m_2 +\nu = \lambda ~.
\end{align}
The conditions (\ref{relat}), (\ref{relat0}) look quite natural from the point of view
of RFT (equality of three-pomeron vertices).
\begin{figure}[h]
\begin{center}
\begin{picture}(300,190)(0,0)
\BCirc(45,190){2}
\BCirc(55,190){2}
\SetWidth{1.5}
\Line(45,188)(45,170)\Line(45,170)(48,160)\Line(48,160)(55,145)
\Line(48,160)(45,145)\Line(45,145)(30,92)
\Line(55,188)(55,165)\Line(55,165)(60,160)\Line(60,160)(55,145)
\Line(60,160)(70,150)\Line(70,150)(65,130)
\Line(55,145)(65,130)\Line(65,130)(70,92)
\Vertex(45,170){1.5}\Vertex(48,160){2}\Vertex(45,145){1.5}
\Vertex(55,145){2}\Vertex(60,160){2}\Vertex(55,165){1.5}
\Vertex(70,150){1.5}\Vertex(65,130){2}
\SetWidth{0.5}
\Line(45,170)(30,140)\Line(30,140)(15,115)\Line(30,140)(40,95)
\Line(15,115)(10,95)\Line(15,115)(20,95)
\Line(55,165)(35,95)\Line(45,145)(50,95)
\Line(70,150)(80,135)\Line(80,135)(60,95)\Line(80,135)(85,95)
\Vertex(30,140){1}\Vertex(15,115){1}\Vertex(80,135){1}
\DashLine(5,90)(90,90){4}\BCirc(30,90){2}\BCirc(70,90){2}
\BCirc(40,10){2}
\BCirc(50,10){2}
\BCirc(60,10){2}
\SetWidth{1.5}
\Line(50,11)(50,30)
\Line(50,30)(37,58)\Line(37,58)(30,88)
\Line(50,30)(58,55)\Line(58,55)(70,88)
\Vertex(50,30){2}\Vertex(37,58){1.5}\Vertex(58,55){1.5}
\SetWidth{0.5}
\Line(40,11)(40,30)\Line(40,30)(20,65)\Line(40,30)(50,60)
\Line(20,65)(15,85)\Line(20,65)(25,85)
\Line(60,11)(60,30)\Line(60,30)(50,60)\Line(60,30)(65,60)
\Line(65,60)(65,60)\Line(65,60)(65,60)
\Line(37,58)(38,85)\Line(50,60)(48,85)\Line(65,60)(58,85)
\Line(65,60)(80,85)\Line(58,55)(53,85)
\Vertex(40,30){1}\Vertex(20,65){1}\Vertex(50,60){1}
\Vertex(65,60){1}\Vertex(60,30){1}
\Text(50,-5)[cc]{$(a)$}
\SetWidth{1.0}
\Vertex(200,40){2}\Vertex(190,150){2}\Vertex(210,150){2}
\PhotonArc(207,97)(28,126,253){2}{5.5}
\PhotonArc(193,97)(28,287,54){2}{5.5}
\Photon(190,120)(200,100){1}{3}\Photon(210,120)(200,100){-1}{3}
\Photon(200,100)(215,84){1}{3}
\Vertex(200,70){2}\Vertex(190,120){2}\Vertex(210,120){2}
\Vertex(200,100){2}\Vertex(215,84){2}
\Photon(200,40)(200,70){2}{3}\Photon(190,150)(190,120){2}{3}\Photon(210,150)(210,120){2}{3}
\Text(200,-5)[cc]{$(b)$}
\end{picture}
\end{center}
\caption{Correspondence between parton cascade history and reggeon diagrams}
\end{figure}
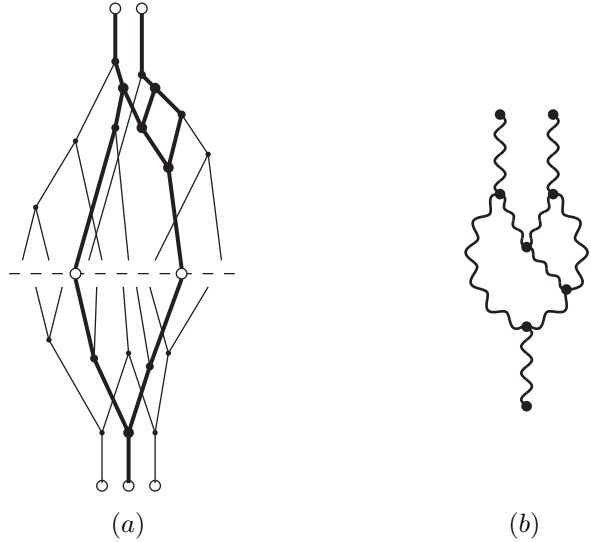

Let us discuss briefly relation between diagrams representing a history of development
of parton cascades and the RFT diagrams. Fig.2a gives an example of evolution
of parton distributions for the colliding hadrons with rapidity. In this example
two partons from each hadron are linked in some intermediate Lorentz frame
($s=2$ in eqs (\ref{T_f}), (\ref{T_g})).
The cascade branches with ``dead ends'', i.e. terminated in noninteracting partons
(thin lines in fig.2a), result in renormalization of distribution functions
(see sect.4) after summation over all possibilities of the ``dead-end'' branches.
The renormalized propagators are showed by thick lines at the figure and form
a {\em skeleton} of the parton network. Each skeleton corresponds to a particular
order of reggeon perturbation theory in number of splittings and fusions of renormalized
reggeon propagators. Different ways of linkage of partons from interacting hadrons
give different skeletons, i.e. different reggeon diagrams.
Skeleton can be imaginated pictorially as a set of lines in the cascade network
with non-zero electric current flow if to ascribe definite potentials to the sets
of initial points of the projectile and target.

Note that the interaction operator (\ref{T_f}) doesn't have a sense of probability
of interaction for two parton systems because it takes into account only single interactions
of partons.
Similar situation one can see in description of nucleus-nucleus scattering in so called
Czy\'{z}-Maximon approximation \cite{C-M} which includes only first-order interactions
of every nucleon. It describes correctly the scattering amplitude only for low nuclear densities,
and account for all orders of nucleon interactions is necessary to get unitary amplitude
for high densities.
However, in case of parton model, introduction of higher order terms into interaction operator
would violate Lorentz invariance of the amplitude.
\section{Examples}

\subsection*{6.1 Eikonal approximation}

Consider first hadron scattering in the model including parton diffusion but
without any parton transformations ($m_1 = m_2 =\lambda = \nu = 0$).
In this approximation each parton walks randomly and independently in the transverse plane,
therefore, if there were no correlations in the initial parton distributions for both
hadrons there will be no correlations in the course of further evolution:
\begin{align}
f_s (y; \Z_s) = \prod_{a=1}^s f_1 (y;z_a) ~, \qquad
\f_s (\y; \tilde{\Z}_s) = \prod_{a=1}^s \f_1 (\y;\tilde{z}_a) ~.
\end{align}
The amplitude of hadron interaction at the impact parameter $b$ is determined
by eq.(\ref{T_f}) and coincides with the eikonal approximation formula
(see fig.3a)
\begin{align}
T^{(eik)}(Y,b) = \sum_{s=1}^{\infty} {(-1)^{s-1}\over s!} \chi (Y,\,b) =
1 - e^{- \chi (Y,\,b)}~,
\end{align}
where $Y=y+\y$, and so called eikonal function is equal to
\begin{align}
\chi (Y,b) = \epsilon \int d\z f_1 (y,\z)  \f_1 (\y,\z-\b) =
\epsilon \int \frac{d\q}{(2\pi)^2}e^{i\q\b} g(y,\q)\g (\y,\q)~.
\end{align}
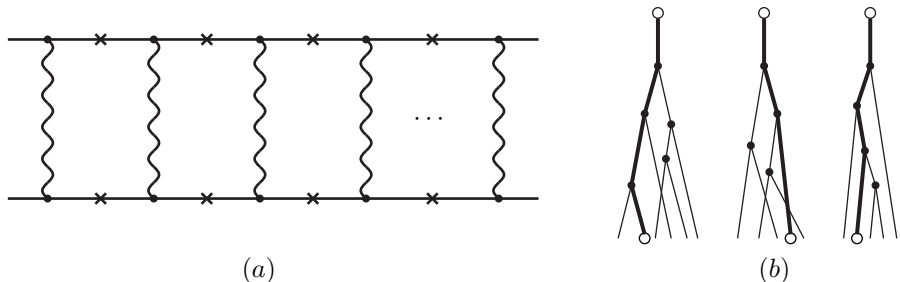
\begin{figure}[h]
\begin{center}
\begin{picture}(300,100)(15,0)
\SetWidth{1}
\Line(-5,80)(195,80)
\Vertex(10,80){1.5}\Vertex(50,80){1.5}\Vertex(90,80){1.5}\Vertex(130,80){1.5}\Vertex(180,80){1.5}
\Line(-5,20)(195,20)
\Vertex(10,20){1.5}\Vertex(50,20){1.5}\Vertex(90,20){1.5}\Vertex(130,20){1.5}\Vertex(180,20){1.5}
\Photon(10,80)(10,20){2}{5}\Photon(50,80)(50,20){2}{5}\Photon(90,80)(90,20){2}{5}
\Photon(130,80)(130,20){2}{5}\Photon(180,80)(180,20){2}{5}
\Line(28,78)(32,82)\Line(68,78)(72,82)\Line(108,78)(112,82)\Line(153,78)(157,82)
\Line(28,82)(32,78)\Line(68,82)(72,78)\Line(108,82)(112,78)\Line(153,82)(157,78)
\Line(28,18)(32,22)\Line(68,18)(72,22)\Line(108,18)(112,22)\Line(153,18)(157,22)
\Line(28,22)(32,18)\Line(68,22)(72,18)\Line(108,22)(112,18)\Line(153,22)(157,18)
\Text(155,50)[cc]{$\cdots$}
\Text(90,-7)[cc]{$(a)$}
\SetWidth{0.5}
\BCirc(240,90){2}\BCirc(280,90){2}\BCirc(320,90){2}
\Vertex(240,70){1.5}\Vertex(280,70){1.5}\Vertex(320,70){1.5}
\Vertex(235,52){1.5}\Vertex(245,48){1.5}\Line(240,70)(245,48)
\Vertex(230,25){1.5}\Vertex(243,35){1.5}\Line(235,52)(245,5)
\Line(245,48)(243,35)\Line(245,48)(255,5)\Line(243,35)(251,5)\Line(243,35)(239,5)
\Line(230,25)(225,5)
\GCirc(235,5){2}{1}
\Vertex(285,52){1.5}\Vertex(275,40){1.5}\Vertex(282,30){1.5}
\Line(280,70)(275,40)\Line(285,52)(282,30)
\Line(275,40)(270,5)\Line(275,40)(285,5)
\Line(282,30)(278,5)\Line(282,30)(295,5)
\GCirc(290,5){2}{1}
\Vertex(315,55){1.5}\Vertex(318,38){1.5}\Vertex(322,25){1.5}
\Line(320,70)(330,5)\Line(315,55)(310,5)
\Line(318,38)(322,25)\Line(322,25)(320,5)\Line(322,25)(325,5)
\GCirc(315,5){2}{1}
\SetWidth{1.5}
\Line(240,88)(240,70)\Line(280,88)(280,70)\Line(320,88)(320,70)
\Line(240,70)(235,52)\Line(235,52)(230,25)\Line(230,25)(235,6.5)
\Line(280,70)(285,52)\Line(285,52)(290,6.5)
\Line(320,70)(315,55)\Line(315,55)(318,38)\Line(318,38)(315,6.5)
\Text(285,-7)[cc]{$(b)$}
\end{picture}
\end{center}
\caption{(a) Reggeon diagrams corresponding to the eikonal approximation;
(b) Parton interpretation of the eikonal approximation for supercritical pomeron}
\end{figure}

If the initial distributions in hadrons were of the $\delta$-function type, then
one-particle distribution at finite $y$ is determined by random walk only:
\begin{align}
f_1 (y,z) = {1\over 4\pi y D} e^{- \textstyle{z^2 \over 4 y D}} ~, ~~~
g_1 (y,q) = e^{- y D q^2 } ~,
\end{align}
and similarly for the hadron $\tilde{h}$. This gives
\begin{align}
\chi (Y,b) = {\epsilon\over 4\pi Y D} e^{- \textstyle{b^2 \over 4 Y D}} ~.
\end{align}
If the initial distribution is given by some function $r(z)$ with Fourier transform
\begin{align}
s(q) = \int\! d\z e^{i\q\z} r(z) ~,
\end{align}
then the eikonal is determined by formula
\begin{align}\label{born}
\chi (Y,b) = \epsilon\int\! {d\q\over (2\pi)^2}\, e^{i\q\b} s(q) \s(q) e^{- Y D \q^2} ~.
\end{align}
It is seen that eq.(\ref{born}) corresponds to exchange of the regge pole with
intercept equaled to $1$, trajectory slope $\alpha^{\prime}=D$ and residue
$s(q) \s(q)$.

Often the formulae of the eikonal approximation are applied in the case  of
supercritical pomeron with intercept larger than $1$. It seems to be inconsistent
in the model under discussion because in the parton picture the supercritical regime
means the growth of parton number when rapidity changes and this can occur only
due to parton splitting in the course of evolution, i.e. at nonzero value of the
constant $\lambda$. The eikonal approximation in the presence of splitting means
that among soft partons originated due to evolution of each initial partons
only one may interact with a target (see fig.3b). The more consistent consideration
of parton splitting allowing interaction with the target for any parton is carried out
in the next example. However, the analytic formulae can be obtained only
if parton diffusion is disregarded.

\subsection*{6.2 The $0$-dimensional Schwimmer model}

The Schwimmer model \cite{Schwimmer75} was developed for description of inelastic
interaction of fast hadron with heavy nucleus and corresponds to summing
enhanced reggeon diagrams in the tree approximation (see fig.4). If a radius of the nucleus
is large enough ($R_A\gg \alpha'Y$), then parton diffusion can be disregarded
and the problem is equivalent to the $0$-dimensional model of sect.2a (we assume that
$\lambda\ne 0$, $m_1\ne0$, $m_2=\nu=0$).\\
\begin{figure}[h]
\begin{center}
\begin{picture}(200,150)(0,10)
\SetWidth{1}
\Line(5,150)(195,150)
\Vertex(100,150){1.5}\Photon(100,150)(100,120){2}{2}\Vertex(100,120){1.5}
\Photon(100,120)(80,80){2}{3}\Vertex(80,80){1.5}
\Photon(100,120)(115,100){2}{2}\Vertex(115,100){1.5}
\Photon(115,100)(100,70){2}{2}\Vertex(100,70){1.5}\Photon(115,100)(160,20){2}{6}
\Photon(100,70)(90,12){2}{4}\Photon(100,70)(120,8){2}{5}
\Photon(80,80)(40,4){2}{6}\Photon(80,80)(80,24){2}{4}
\Line(5,4)(195,4)\Line(5,8)(195,8)\Line(5,12)(195,12)\Line(5,16)(195,16)\Line(5,20)(195,20)
\Line(5,24)(195,24)
\end{picture}
\end{center}
\caption{Reggeon diagrams corresponding to the Schwimmer approximation}
\end{figure}
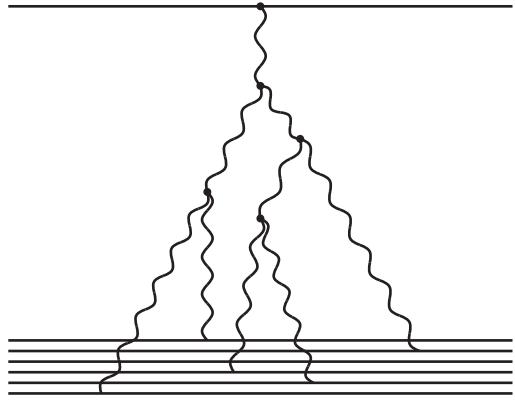
Choose the initial state of the projectile hadron as containing one parton with
a probability $g_p$
\begin{align}\label{Sch-p}
& p_1^{(p)} = g_p ~,~~p_0^{(p)} = 1-g_p ~, ~~p_N^{(p)} = 0 ~\text{~~at ~}N\ge 2 ~, \nn\\
& G^{(p)}(w,0)=1 + g_p (w-1) ~,
\end{align}
and the initial parton distribution for the target in the Poisson form
\begin{align}\label{Sch-t}
& p_N^{(t)}(0) = {g_t^N \over N!} e^{-g_t} ~, ~~ G^{(t)}(w,0)= e^{g_t(w-1)} ~,
~~ \mu_s^{(t)}(0) = (g_t)^s.
\end{align}
The fusion is absent in this model, $\nu =0$, so the Lorentz-invariant consideration
is impossible. The interaction amplitude calculated in the laboratory frame
has according to the 0-dimensional version of the formula (\ref{T_f})
the following form:
\begin{align}
T_{schw}(Y) &= \sum_{s=1}^{\infty} {(-1)^{s-1}\over s!} \mu_s^{(p)}(Y)\;\mu_s^{(t)}(0)
\nn\\ &= 1 - G^{(p)}(1-g_t;Y)
={g_p g_t e^{\Delta Y}
\over 1+{\ds g_t \lambda \over \Delta} (e^{\Delta Y} -1)}~.
\end{align}
We used the definition (\ref{G0}) for the generating function $G^{(p)}$ and its explicit form
(\ref{G0_1}), (\ref{G0_2}) as applied to the initial condition (\ref{Sch-p}):
$G^{(p)} (w;Y) = g_p W(Y) $, with $W(Y)$ from eq.(\ref{G0_2}).

When one estimates the cross section of hadron interaction with a nucleus $A$ one has
to choose $g_t$ to be proportional to the nuclear density $\rho_A (\b)$:
\begin{align}
g_t = g_\N \,\rho_\A (\b) ~.
\end{align}
For heavy nucleus ($A\gg 1$) in the approximation of constant density,
\begin{align}
\rho_\A = \frac{A}{\pi R_\A^2} \theta (R_\A - |\b|) ~, ~~~ R_\A = r_0 A^{1/3} ~,
\end{align}
we come to
\begin{align}\label{schwimmer}
T_\A (Y,b) = \frac{g_p\, g_\N A^{1/3}}{\pi r_0^2}~{\exp (\Delta Y) \; \theta (R_\A-|\b|)
\over 1 + {\ds \lambda g_\N A^{1/3}\over \ds r_0^2}
\left({\ds \exp (\Delta Y) -1 \over \ds \Delta }\right)}
\end{align}
This formula coincides with the original Schwimmer formula for amplitude of
hadron--nucleus scattering in the approximation of very large nucleus.
When $Y\rightarrow\infty$ the amplitude goes to the constant limit
$g_p \,\Delta / \lambda = g_p (1- m_1 / \lambda) < 1$. Note that in general
reggeon field theory the constants $g_p$, $\Delta$ and $\lambda$ may be arbitrary.

It is worth to stress that though the formula (\ref{schwimmer}) coincides in form
with the Schwimmer formula obtained in RFT, the meaning
of the constants entering these formulae is different, and even their dimensions
differ: in eq.(\ref{schwimmer}) the constants $g_p$ and $\lambda$ are dimensionless,
and $g_\N$ has dimension of length squared, while in reggeon field theory
all these constants have dimension of length.

It is easy to formulate a generalization of the Schwimmer approximation in this approach.
If the initial parton distribution in the projectile has the Poisson form, i.e.
\begin{align}
\mu_s^{(p)}(0) = (g_p)^s ~, ~~ G^{(p)} (w,0) = \exp [g_p w ]~, ~~
G^{(p)} (w,Y) = \exp [g_p \w (Y)] ~,
\end{align}
one gets the ``eikonalized'' Schwimmer formula
\begin{align}\label{eik_schw}
T_{schw}^{(eik)}(Y) = 1 - \exp [T_{schw} (Y)] ~.
\end{align}

Note that analysis of interaction of {\em two} heavy nuclei requires even in the tree approximation
taking into consideration not only parton splitting but their fusion also.
Moreover, as it was emphasized already, the Lorentz-invariant parton consideration
is possible only in presence both splitting and fusion of partons provided
the corresponding constants are related.

\subsection*{6.3 Asymptotics of the $0$-dimensional model with account for parton fusion}

We consider in this section the $0$-dimensional model including splitting and fusion of
partons for the case $m_1 = m_2 =0$.
Let us demonstrate the Lorentz invariance of the parton picture for asymptotically large
rapidity value $Y$, when a stationary regime sets in and the interaction amplitude does not
depend on $Y$ at all.
Indeed, in this case for any frame choice one of hadron rapidities (say, $y$) is large.
Hence the projectile hadron is in asymptotical stationary state (see sect.2.b)
\begin{align}
\mu_s^{(p)} (y) = B^{(p)} a^s ~, ~~~
B^{(p)}={1-G_0^{(p)}(0)\over 1-e^{- a}}~, ~~~
a = \lambda/\nu = 1 ~,
\end{align}
and the state of a target with rapidity $\y$ is given by the generating function $G^{(t)}(\y,w)$.
It is essential that the value of this function at $w=0$ does not depend on $\y$
and is equal to $G_0^{(t)}(0)$.

Therefore for $a =1$ the interaction amplitude does not depend on rapidity in any
Lorentz frame and equals
\begin{align}
T_{\infty}(Y) = \sum_{s=1}^{\infty} {(-1)^{s-1}\over s!} \mu_s^{(p)}(y) \mu_s^{(t)} (\y)
&=B^{(p)}\left[ 1 - G^{(t)}(\y, 1-a) \right] \nn\\
&={[1-G_0^{(p)}(0)] [1-G_0^{(t)}(0)] \over 1-e^{-1}} ~.
\end{align}

\subsection*{5. Two-dimensional diffusion}

In presence of parton diffusion it is possible to distinguish in the parton structure
of the fast hadron two characteristic regions.
At the periphery where parton density is small the parton fusion is insignificant and only
diffusion and splitting processes occur.
Parton distribution in this region behaves like the Green function of diffusion equation
complemented with exponential increase due to splitting:
\begin{align}\label{periph}
f_1^{(periph)}(y,z) = e^{\lambda y}\frac{1}{4\pi y}e^{-z^2/4Dy} ~.
\end{align}
In the central region the parton fusion is essential and the density comes to the
stationary value
\begin{align}\label{centr}
f_1^{(centr)}(y,z)\approx\lambda/\nu ~.
\end{align}
Transition from one regime to another corresponds to equality of the central and
peripheral density values, i.e. to distances of order
\begin{align}\label{range}
R(y)\approx 2\sqrt{\lambda D}y ~.
\end{align}
At larger distances the density decreases exponentially. Thus, the fast hadron represents
to be a disc of constant density of radius increasing proportionally to its rapidity:
\begin{align}\label{asympt}
 f_1(y,z)\simeq{\lambda\over\nu} \;\theta [R(y)-z] ~.
\end{align}
Obviously the interaction cross section of two such discs corresponds to the Froissart
regime: $\sigma (Y) \approx \text{const }\cdot Y^2$, where $Y=y+\y$. In order to calculate
the constant value, i.e. the degree of blackness, it is necessary to account for
contributions of all multi-particle densities $f_s(y,\Z_s)$, $\f_s(\y,\Z_s)$.
It is not the aim of this paper to carry out complete analysis of this problem.
Let us consider as an illustration only the simplest case when the initial state
of one of hadrons (say, $\h$) contains only one parton, that is
$\f_1(\y=0,z)=\delta(b-z)$, $\f_s(\y=0,\Z_s)=0, s\geq 2$.
Let us use the Lorentz invariance of the problem and apply the formula (\ref{T_f})
in the rest frame of the hadron $\h$. Then in the formula only the first term
remains:
\begin{align}\label{disk1}
T(y,\y;b)
&=\,\epsilon \int\! dz  f_1 (Y;z) \f_1 (0;\tilde{z}) \delta(\z-\tilde{\z}-\b) \nn\\
&\approx \frac{\epsilon\lambda}{\nu}\,\theta (R(Y)-b) = \theta (R(Y)-b) ~,
\end{align}
which corresponds due to relation (\ref{relat}) to scattering on the absolutely
black disk of radius $R(Y)$.

For more complicated cases (many-parton initial states) the situations turnes out to be
the same --- the asymtotics of the amplitude corresponds to the  scattering off a
black disc.
In ref. \cite{kancheli00} Kancheli discussed the problems related to the Lorentz invariance
of interaction of parton systems for the case of the {\em gray} disc (with finite transparency).
In present model this problem does not appear, and both Lorentz invariance and blackness
of the amplitude are provided by relation (\ref{relat}).
\section{Unitarity}

One of the key problems of RFT is $s$-channel unitarity of the scattering amplitude.
As can be seen, the unitarity of the amplitude calculated for the model under consideration
depends not only from the parton dynamics but also from the initial conditions,
i.e. from the form of the parton sources for each hadron.

Let us discuss under which conditions the amplitude $T(Y,b)$ can be interpreted
as a probability of interaction at fixed impact parameter that is
satisfies to the relation
\begin{align}\label{unit}
0 < T(Y,b) < 1 ~.
\end{align}

Consider first the case with $d=0$ and note that if both sets of moments
$\mu_s(y)$, $\tilde{\mu}_s(Y-y)$ in equation
\begin{align}\label{int_0}
T(Y,b)= \sum_{s=1}^{\infty} {(-1)^{s-1}\over s!} \mu_s(y)\tilde{\mu}_s(Y-y)
\end{align}
would be of Poisson form,
\begin{align}\label{Poiss}
\mu_s=\alpha^s ~, \qquad \tilde{\mu}_s = \tilde{\alpha}^s ~,
\end{align}
the amplitude has a simple form satisfying to relations (\ref{unit}):
\begin{align}\label{Poiss_int}
T^{(\alpha,\tilde{\alpha})}(Y,b)= 1 - \exp (\alpha\tilde{\alpha}) ~.
\end{align}

This observation makes useful so called Poisson representation \cite{Gardiner}
for probabilities.
It represents (for the simplest case) probabilities $p_N (y)$ as superpositions
of Poisson distributions $P_N (\alpha)$:
\begin{align}\label{Poiss rep}
p_N (y) = \int d\alpha f(\alpha, y)\left\{ {\alpha^N\over N!}e^{-\alpha}\right\}
=\langle P_N (\alpha)\rangle ~,
\end{align}
where we used notation $\langle\dots\rangle$ for averaging with weight $f(\alpha,y)$.

The generating function $G(w;y)$ can be written as
\begin{align}\label{G_f}
G(w;y) = \int d\alpha\,f(\alpha, y)\, {\exp[(w-1)\alpha]}  ~.
\end{align}
The master equation (\ref{gen_0}) gives for $f(\alpha,y)$ the equation of
Fokker-Planck type
\begin{align}\label{f}
\frac{\d f(\alpha, y)}{\d y} = -\frac{\d}{\d\alpha}\left\{ A(\alpha)\right\} +
\frac{1}{2} \frac{\d^2}{\d\alpha^2}\left\{ B(\alpha) \right\} ~.
\end{align}
where functions
\begin{align}\label{A,B}
A(\alpha) = (\lambda-m_1)\alpha - (\nu + 2m_2)\alpha^2 ~, \qquad
B(\alpha) = 2\lambda\alpha - 2(\nu+m_2)\alpha^2 ~,
\end{align}
describe driving term and diffusion coefficient correspondingly. The diffusion
coefficient $B$ is non-negative for $0\le \alpha\le \lambda/(\nu+m_2)$.
The condition (\ref{bound1}) provides normalization of $f(\alpha, y)$:
\begin{align}\label{norm_f}
\int d\alpha f(\alpha, y) = 1 ~.
\end{align}
Equation (\ref{f}) is equivalent \cite{Gardiner} to the differential
{\em stochastic} equation
\begin{align}\label{f_stoch}
d\alpha = A(\alpha) dy + \sqrt{B(\alpha)} dW(y) ~,
\end{align}
where $dW(y)$ is a stochastic differential corresponding to the {\em Wiener process}.

Suppose for simplicity that $m_1=m_2=0$. One can see from (\ref{f_stoch}) that if
evolution starts at $y=0$ at $\alpha$ value {\em within} interval $(0,a)$,
$a=\lambda/\nu$, then the first term will drive the system to the right end of the interval
$a=\lambda/\nu)$. At this point both the driving term and diffusion coefficient vanish
and, therefore, it is a stationary point. Any initial $\alpha$-distribution will be
concentrated at this point when $y\rightarrow \infty$. Thus the asymptotic distribution
is Poisson distribution $P_N(a)$, in correspondence with analysis of section 2.

Because of positivity of the diffusion coefficient $B(\alpha)$ at the interval $(0,a)$,
the distribution $f(\alpha, y)$ is positive and normalizable in the course of
evolution, i.e. it can be considered as a probability distribution.
If both $f(\alpha, y)$ and $\f(\alpha, \y)$ are of class of probability
distributions, then the interaction amplitude $T(Y,b)$ is given by averaging
of equation (\ref{Poiss_int}) over $\alpha$ and $\tilde{\alpha}$,
\begin{align}
T(Y,b) = \langle\langle T^{(\alpha,\tilde{\alpha})}(Y,b)\rangle\rangle
\equiv\int d\alpha f(\alpha,y) \int d \tilde{\alpha} f(\tilde{\alpha},\y)
\left[1 - \exp (\alpha\tilde{\alpha})\right] ~,
\end{align}
and evidently satisfies the unitarity condition (\ref{unit}).
Consideration of two-dimensional case can be carried out similarly.
\section{Conclusions}
\indent

The parton stochastic model discussed in this paper is in exact correspondence with reggeon
field theory. Only special reggeon field theories with particular set of pomeron vertices
allow parton interpretation of this sort.
The model reproduces results of main approximations for RFT.
In the theory with zero number of transverse dimensions, $d=0$, it is possible to analyze
quantitatively an asymptotical regime. It is equivalent to summing all diagrams of 0-dimensional
RFT with all loops. For theory with $d=2$ qualitative and numerical consideration is possible.
Because of its classical and stochastic nature the model allows numerical simulation by means
of simple Monte Carlo algorithm.
Direct analogy with kinetic theory of chemical reactions gives good prospects for application
of thermodynamical methods and of theory of stochastic equations.

The model allows a number of generalizations, in particular to parton systems with internal
quantum numbers and of various spacial scales.
\section*{Acknowledgments}
\addcontentsline{toc}{section}{\numberline{}Acknowledgments}
The author is thankful to A.B.~Kaidalov and O.V.~Kancheli for
numerous stimulating discussions and to M.A.~Braun,
Yu.L.~Dokshitzer and Yu.A.~Simonov for discussion of results. This
work has been funded in part by the Russian Foundation of Basic
Researches (RFBR 98-02-17463, 00-15-9678), by the NATO grant PSTCLG 977275
and by the INTAS grant 00-00366.
\section*{References}
\addcontentsline{toc}{section}{\numberline{}References}

\end{document}